\documentstyle[preprint,aps,prb]{revtex}
\begin{document}
\draft



\title{Magnetic field dependence
of the superconducting fluctuation contribution
to NMR-NQR relaxation}



\author{P. Mosconi, A. Rigamonti and A. A. Varlamov\footnote{On leave of absence from
Department of Theoretical Physics, Moscow Institute for Steel and Alloys,
 Leninski pr. 4, Moscow 117936, Russia}}
\address{Department of Physics \lq\lq A. Volta\rq\rq, Unit\`a INFM and 
Sezione INFN,
Via Bassi 6, I-27100 Pavia, Italy}


\maketitle



\begin{abstract}
The dependence of the Maki-Thompson
(MT) and of the density of states depletion (DOS)
contributions from 
superconducting fluctuations
(SF) to NMR-NQR relaxation is derived
in the framework of the
diagrammatic theory,
applied to layered three dimensional
(3D) high $T_c$ superconductors.
The regularization procedure
devised for
the conductivity (Buzdin and Varlamov,
Phys. Rev. B, 58, 14195 (1998))
is used in order to avoid the
divergence of the DOS term.
The theoretical results are 
discussed in the light
of NMR-NQR
measurements in YBCO
and compared with
the recent theory
(Eschrig $et$ $al.$, Phys. Rev. B 59, 12095 (1999)),
based on the assumption of
a purely 2D spectrum of fluctuations.




\end{abstract}



\newpage
\section{ INTRODUCTION}
The normal state
of high temperature
superconductors
(HT$_c$SC)
is characterized by unusual
properties, most of them
still lacking of a comprehensive
theoretical description.
In particular,
the transfer of spin excitations from 
 low to high-energy
range (spin-gap opening)
well above $T_c$,
detected in the generalized
susceptibility
from $T_1$
and neutron scattering measurements
(for a recent review see Ref. 1),
and the related quasi-particle
gap, observed in ARPES$^{2,3}$,
have been tentatively
related to superconducting fluctuations
(SF) of various nature.
Among them, one could
mention preformed pairs 
without long-range
phase coherence, possibly along stripes$^{4,5}$,
spin and/or charge density waves$^{6,7}$,
coupling of a d-wave symmetry order parameter
with spin fluctuations$^{8,9}$,
order parameter fluctuations 
well beyond the 
perturbative approach$^{10-12}$
and quantum critical point
fluctuations$^{13}$.
For a review on precursor
pairing correlations
and a survey of the various scenarios
see Ref. 14.
Furthermore the magnetic
field has been argued$^{15}$ to have a role
on SF of overdoped HT$_c$SC also,
by inducing
a spin-gap from
$T_c(0)$ to $T_c(H)$.


The role
of the magnetic field is crucial in NMR-NQR
attempts to study SF in the vicinity
 of $T_{c}^{+}$.
Indeed, the most
direct contribution
to SF, namely the 
Aslamazov-Larkin term,
responsible of paraconductivity$^{12}$,
is not effective
in causing an extra-contribution to 
 NMR-NQR relaxation.
In principle, the SF contribution
 to the relaxation
rates, for $T\to T_{c}^{+}$,
are the Maki-Thompson (MT) one
(related to the pairing of a carrier
with itself
at a previous stage of motion),
and the reduction due to
the depletion in the single-particle 
density of states (DOS),
when fluctuating pairs are
created$^{12}$.
These two terms might
have a different sensitivity to
the presence
of a magnetic field,
which acts as a pair-breaking factor.


The first NMR experimetal
observation$^{16}$ of the role
played by  SF in HT$_c$SC
was based on the comparison
of $^{63}$Cu
relaxation rate $W$ in YBCO in the 
absence (i.e. NQR)
and in the presence of a magnetic field
of about 6 T.
Within about 10 degrees above $T_{c}(0)$,
$W(NQR)$ was found
to decrease upon application
of the field 
by a factor about 5$\div$15\%.
A qualitative interpretation
of the experimental observation
was given by assuming
that the field reduced
the MT term to about 25\%,
while the more
robust
DOS term was little
affected by the field.
The Equations used in
 these estimates$^{16}$
were the ones$^{17}$ pertaining
to a three-dimensional (3D)
layered spectrum of
excitations
(with anisotropy parameter
$r=2\xi_{c}^{2}(0)/d^{2}\simeq 0.1$,
where $\xi_c(0)$ is the correlation
lenght of the Cooper pair
at zero temperature, along the $\hat c$-axis, and
$d$ the interlayer distance).
The occurence of a pure 2D regime of SF
could be ruled out,
on the basis of the absolute
value and of the temperature
dependence of the SF contribution to
$^{63}$Cu.


The first systematic
analysis of the field
dependence
of NMR relaxation rate 
in YBCO was carried out
in 1998 by Mitrovi\'c $et$ $al.^{18}$,
by varying the field from 
2.1 up to 27.3 T.
The $^{63}$Cu $T_1$ was
probed$^{18}$ through
the contribution
to the $^{17}$O echo dephasing.
For field in the range 6$\div$8 T,
the results derived in this way$^{18}$
were found to coincide
with the direct measurements of $^{63}$Cu
$T_1$.
At high field, $W^{DOS}$
was argued to be strongly reduced
by the field.
These data$^{18}$, as well as the field dependence
of $^{17}$O(2,3) Knight shift$^{19}$,
have been interpreted on the basis 
of a theory for the DOS
contribution due to Eschrig $et$ $al.^{20}$,
which extended analytical
approaches$^{17}$ to include short wave-lenght
and dynamical fluctuations,
in the assumption
of a 2D regime. 


Recently, Gorny $et$ $al.^{21}$ 
reported precise $^{63}$Cu relaxation
measurements
in YBCO for $H=0$, 8.8 and 14.8 T,
finding no magnetic
field dependence in a wide temperature
range.
A possible dependence
of the field effect
on the amount of doping
could be suspected since,
 at the same time,
no field effect had been
observed$^{22}$ in underdoped YBCO.


In this paper we derive
the magnetic field dependence
of the MT and DOS contributions
to NMR $T_1$, in the framework
of a diagrammatic
description,
for arbitrary values of the reduced
field $\beta=2H/H_{c2}(0)$,
for a 3D layered spectrum of fluctuations, which should
pertain to the
case of YBCO with low anisotropy
parameter.
In order to remove the
logarithmic
divergence
present in the DOS term, here
we use the method devised$^{23}$ 
for transverse conductivity, in
which 
regularization requirement,
analogous to the ones for the
nuclear relaxation rate, is present.
Furthermore,
we briefly discuss the role
of the long wave-vector
fluctuations and of the dynamical 
fluctuations.
Our analytical conclusive expressions
are compared
with the numerical
solutions$^{20}$
for the 2D regime and with
the experimental measurements
carried out until now.









\section{Field dependence of the SF contribution
nuclear relaxation}


In the following, we extend
 the diagrammatic
theory for the SF contribution
to NMR-NQR
relaxation rate,
to include the effects due to the 
presence of a magnetic field along the $\hat c$-axis,
in HT$_c$SC.


In the presence of the field
the MT and DOS contributions to
the relaxation rates$^{12,17}$
must be evaluated 
by starting from
the usual inclusion in the momentum ${\bf q}$
of the term $(-2e/c){\bf A}$,
with ${\bf A}=\frac{B}{2}(-y{\bf i}+x{\bf j})$.
The integration over ${\bf q}$,
in the $ab$ plane,
is substituted by a sum over the Landau levels.
Thus the MT and DOS contributions$^{17}$ to $W$
assume the forms


\begin{eqnarray}
\frac{W^{MT}}{W^{0}}(\beta,
\varepsilon)&=&\frac{\pi}{8E_F\tau}\frac
{\beta}{(\varepsilon -\gamma
_\varphi )}
\cdot\sum_n
\biggl[\frac 1{\sqrt{\gamma _\varphi +\beta
(n+1/2)}\sqrt{\gamma _\varphi +\beta (n+1/2)+r}}+\nonumber\\ 
&&-\frac 1{\sqrt{\varepsilon +\beta (n+1/2)}\sqrt{\varepsilon +\beta
(n+1/2)+r}}\biggr]
\label{MTgen}
\end{eqnarray}
and
\begin{eqnarray}
\frac{W^{DOS}}{W^0}
({\beta },\epsilon )&=& 
-\frac{\hbar}{E_F \tau}\cdot\kappa(T\tau)
\cdot\sum_n^{1/\beta }\frac 1{\sqrt{\varepsilon +\beta (n+1/2)}\sqrt{%
\varepsilon +\beta (n+1/2)+r}}
\label{DOSgen}
\end{eqnarray}
where $W^0$ is the ordinary
 relaxation rate
in the absence of SF
(the Korringa one in a Fermi gas-like model).
In the above Equations,
$E_F$ is the Fermi energy,
$\tau$ the single particle collision time,
$\gamma_{\varphi}=\xi_{0}^{2}/{\bf D}\tau_\varphi$
(with 
 ${\bf D}=E_F\tau /m$, 2D carrier diffusion constant
and $\hbar\tau_{\varphi}^{-1}$ depairing energy)
is a dimensionless factor
which, in the limit $B\to 0$,
takes into account the pair-breaking
effect and $\varepsilon=(T-T_c)/T_c$
is the reduced temperature.
$\kappa(T\tau)$ in Eq. (2) is the function



\[
\kappa (T\tau )=\frac{7\zeta (3)}\pi \frac 1{4\pi T\tau \left[ \psi
(1/2)-\psi (1/2+1/4\pi T\tau )\right] +\psi ^{\prime }(1/2)}= 
\cases
{\frac{14\zeta (3)}{\pi ^3},&$T\tau \ll 1$ \cr 
4T\tau ,&$T\tau \gg 1$\cr}
\]
where, $K_B=\hbar=1$ and $\psi$ is the 
Euler digamma function.
The expressions (1) and (2) are general
and hold for any
magnetic field,
provided that
$\beta=2B/B_{c2}(0)\ll 1$ (see Ref. 24).
 The DOS term
shows a logarithmic divergence
in the sum. 


The behaviours of $W^{MT}$ and $W^{DOS}$,
for small fields,
can easily be derived from Eqs. (1) and (2),
by a straightforward
expansion.
The intermediate and strong field
regimes, namely $\varepsilon\ll \beta\ll 1$,
require some care.
For the MT contribution
(Eq. (1)) the sum
over $n$ converges rapidly
and thus, for strong fields,
only the term $n=0$ can be taken into account.
For the DOS contribution (Eq. (2)),
however,
the logarithmic divergence
must be removed.
This can be achieved
by considering the cut-off 
independent difference


\begin{equation}
\Delta W^{DOS}(\beta,\varepsilon)\equiv
W^{DOS}(\beta,\varepsilon)-W^{DOS}(0,\varepsilon)
\end{equation}


The zero field 
expression is rewritten in the form


\begin{eqnarray}
\frac{W^{DOS}({0},\varepsilon )}{W^{0}}=
-\frac{\hbar}{E_F \tau}\cdot 2\kappa(T\tau)
\lim_{\beta \rightarrow 0}
\sum_{n=0}^{1/{\beta }}\ln {\frac{\sqrt{\varepsilon +\beta
n+\beta }+\sqrt{\varepsilon +r+\beta n+\beta }}
{\sqrt{\varepsilon +\beta n}+\sqrt{%
\varepsilon +r+\beta n}}}
\label{zerofield}
\end{eqnarray}
and then, Eq. (3) becomes


\begin{eqnarray}
\frac{\Delta W^{DOS}({\beta },\varepsilon )}
{W^0}&=&
\frac{\hbar}{E_F\tau}\cdot2\kappa(T\tau)
\sum_{n=0}^\infty \biggl\{ \ln 
{\frac{\sqrt{\varepsilon
+\beta n+\beta }+\sqrt{\varepsilon +r+\beta n+\beta }}
{\sqrt{\epsilon +\beta n}+%
\sqrt{\varepsilon +r+\beta n}}}+ \nonumber \\
&&-\frac{\beta} {2\sqrt{\varepsilon +\beta n+\beta /2}
\sqrt{\varepsilon +r+\beta n+\beta/2}}\biggr\}
\label{delta2}
\end{eqnarray}
where the summation
has been extended up to $n\to \infty$,
since 
$\Delta W^{DOS}(\beta, \varepsilon)$
now involves a summation
which is convergent
(the n-th term being proportional to
$n^{-3/2}$ for large $n$).


 Now we are going to
discuss the behaviours of
$W^{DOS}$
and $W^{MT}$ in the asymptotic limits,
corresponding to different
temperature and field regimes.


\subsection{Weak magnetic fields}


In the weak field regime, namely $\beta\ll \varepsilon$,
Eqs. (1), (2) can be expanded in powers of $\beta$.
For the DOS term, in particular,
the Euler-MacLaurin formula


\begin{eqnarray*}
\sum_{n=0}^Nf(n)=\int_0^Nf(x)dx+{\frac 12}[f(N)+f(0)]+{\frac 1{{12}}}%
[f^{^{\prime }}(N)-f^{^{\prime }}(0)]+\ldots 
\end{eqnarray*}
yields
\begin{eqnarray}
\frac{W^{MT}}{W^0}
(\beta\ll \varepsilon)&=&
\frac{\pi}{8E_F \tau}\frac{1}
{\varepsilon-\gamma_{\varphi}}\biggl[2\ln \biggl(\frac{\varepsilon^{1/2}+
(\varepsilon+r)^{1/2}}
{\gamma_{\varphi}^{1/2}+(\gamma_{\varphi}+r)^{1/2}}\biggr)+\nonumber\\
&-&\frac{\beta^2}{24}\biggl(\frac{\gamma_{\varphi}+r/2}
{[\gamma_{\varphi}(\gamma_{\varphi}+r)]^{3/2}}-
\frac{\varepsilon+r/2}
{[\varepsilon(\varepsilon+r)]^{3/2}}\biggr)\biggr]
\label{MTw}
\end{eqnarray}


\begin{eqnarray}
\frac{W^{DOS}}{W^0}
(\beta\ll\varepsilon ) &=&-\frac{\hbar}
{E_F \tau}\cdot \kappa(T\tau)
\biggl[2\ln \biggl({\frac 2{{%
\varepsilon ^{1/2}+(\varepsilon +r)^{1/2}}}}\biggr)+\nonumber\\
&-&{\frac{{\beta ^2(\varepsilon +r/2)}}
{{24[\varepsilon (\varepsilon +r)]^{3/2}}}}%
\biggr]
\label{DOSw}
\end{eqnarray}


\subsection{Intermediate and strong field
regimes}


For the MT term in the strong field
regime ($\beta\gg \max\{\varepsilon,r,\gamma_\varphi\}$),
by expanding Eq. (1) in powers of $\beta^{-1}$,
one has


\begin{equation}
\frac{W^{MT}}{W^0}
(\varepsilon ,\gamma _\varphi ,r\ll \beta )=\frac{%
3\pi ^3}{16E_F\tau}\frac {1}{\beta}
\end{equation} 


In the intermediate case
(namely $\varepsilon,\gamma_\varphi \ll\beta\ll r$
and $\varepsilon\ll\beta\ll\gamma_\varphi, r$),
which can become relevant for a 3D layered compound,
the series expansions 
in terms of the smallest parameters yield


\begin{equation}
\frac{W^{MT}}{W^0}
(\varepsilon ,\gamma _\varphi ,\ll \beta \ll r)=
4.57 \frac{\pi}{16E_F\tau}\frac {1}{\sqrt{\beta r}}
\end{equation}


\begin{equation}
\frac{W^{MT}}{W^0}
(\varepsilon \ll \beta \ll \gamma _\varphi ,r)=%
\frac{\pi}{8E_F\tau}\frac 1{\gamma _\varphi }\ln \frac{\sqrt{\max
\{\gamma _\varphi ,r\}}}{\sqrt{\beta }+\sqrt{\beta +r}}.
\end{equation}


For the field dependence of the
DOS correction to the zero field
contribution (Eq. (5)), in strong field regime,
one can take into account
the $n=0$ term only:


\begin{eqnarray}
\frac{\Delta W^{DOS}
(\beta \gg \max \{\varepsilon ,r\})}
{W^0}=
\frac{\hbar}{E_F\tau}\cdot 2 \kappa(T\tau)
\biggl\{\ln {\frac{2\sqrt{%
\beta }}{e(\sqrt{\varepsilon }+\sqrt{\varepsilon +r})}}\biggr\} 
\label{strongfield}
\end{eqnarray}
The $n\geq 1$ terms, neglected in this evaluation,
yield a correction of 0.02 in the bracket term
in Eq (11).


In the intermediate fields, namely 
for $\varepsilon\ll \beta \ll r$
(corresponding to a 3D layered
regime of fluctuations),
by means of an expansion of Eq. (5)
in terms of $\beta/2r$ one has


\begin{eqnarray}
\frac{\Delta W^{DOS}(\varepsilon \ll \beta \ll r)}
{W^0}=
0.428\frac{\hbar}{E_F\tau}\cdot\kappa(T\tau)
\sqrt{\frac \beta {2r}}.
\label{interfield}
\end{eqnarray}







\section{Discussion and comparison
with experimental results}


As it appears from the theoretical
treatment given in the previous
Section,
the dependence on
the magnetic field of the SF contributions
to the NMR-NQR
relaxation rate is a rather delicate
issue, because of the non-trivial
interplay of several parameters,
such as $(T-T_c)$,
the reduced field $\beta=2B/B_{c2}(0)$,
the anisotropy parameter
$r$, the elastic collision time $\tau$ and
the anelastic phase-breaking time
$\tau_\varphi$. 


Eschrig $et$ $al.^{20}$, in their
numerical extension of the previous 
analytical approaches$^{17}$,
have considered arbitrary values of
$\kappa(T\tau)$
and taken into account
short wave-lenght and dynamical fluctuations.
However, this generalization
has required the restriction to a purely
2D regime of SF,
which seems
questionable in view of
experimental findings
in YBCO, pointing out
a crossover to 3D fluctuations
well above $T_c$,
at least for relatively small 
fields$^{25-27}$.
On the other hand,
it could be remarked that
dynamical fluctuations
are relevant only when the field
is comparable to $B_{c2}(0)\sim 100\div 120$ T.
Therefore the reduction to static
and small wave-vector 
fluctuations should not
invalidate our
conclusive expressions 
given above.


To illustrate the theoretical expressions,
derived in Section II,
we plot in Figs. 1 and 2 the
temperature and field behaviours
of $W^{DOS}$ and $W^{MT}$,
with a choice of parameters
appropriate to YBCO optimally doped.


In Fig. 3 the experimental
results from various
authors (and from different YBCO samples,
about optimally doped) are compared
with the behaviour expected for DOS contribution,
according to  
 Eqs. (3) and (5).
A relatively
small field dependence,
of both DOS and MT contributions,
up to a reduced field of about 0.2,
is noticed.
For strong fields
a reduction of the MT
term can be
expected, while the DOS term
seems only slightly affected (see Figs. 1b and 2b).
As a consequence,
either for s-symmetry
of the fluctuating Cooper
pair (namely, presence
 of both the MT and DOS
 effects of SF on $T_1$)
or for d-symmetry (and therefore,
no MT contribution)
only a slight dependence
of $T_1$, on the magnetic field, should be
detected,
as it is illustrated in Fig. 3,
for the DOS term.
In particular the experimental data$^{18}$
at strong fields ($\beta> 0.2$)
can hardly be justified.
The discrepancy
between the theoretical
description and the experimental
findings, at strong fields, might be related to the
general framework involved in our treatment.
Howerver, it should be
stressed that a breakdown of the Fermi liquid
picture should not invalidate
 the expression 
for the DOS
contribution,
which is independent from
the pairing mechanism and the
normal state properties.
The dimensional crossover (3D$\to$2D)
for strong fields is taken into account
 in our equations, but
it could be argued that
dynamical and short wave-lenght
fluctuations
are no longer negligible 
in a 2D strong field regime.
Finally, it should be taken into
consideration the possibility
that the method of an indirect estimate
of $^{63}$Cu $T_1$ from $^{17}$O
echo dephasing$^{18}$ could
be invalidated in strong fields.


Summarizing,
from the behaviour of $^{63}$Cu
$T_1^{16,18,19,21}$
one can infer that a DOS SF 
contribution to the nuclear 
relaxation is present in the 
vicinity of $T_c$.
The MT term is more elusive,
requiring a s-wave component in the
spectrum of the SF and being
possibly strongly sample-dependent,
through the pair-breaking and
impurities effects.
The field dependence of DOS
term is a rather delicate issue,
complicated by various crossovers
and parameters. Further
theoretical and experimental
work is required for a firm
conclusion.





\section*{Acknowledgments}


P. Carretta is gratefully thanked for
providing some experimental
data before publication and for
useful discussions.
Stimulating discussions with
 W. Halperin, M. Eschrig, V. Mitrovi\'c
and K. Gorny are gratefully acknowledged.


\newpage
\section*{REFERENCES}
\begin{enumerate}


\item  A. Rigamonti, F. Borsa, C. Carretta,
\lq\lq Basic aspects and main results of NMR-NQR
spectroscopies in HTSC\rq\rq, Rep.
Prog. Phys., 61, 1367 (1998)


\item A. G. Loeser 
$et$ $al.$ Science 273, 325 (1996)  


\item H. Ding 
$et$ $al.$ Nature 382, 51 (1996)


\item S. Doniach and M. Inui, Phys. Rev. B 41, 6668 (1990)


\item V.J. Emery and S.A. Kivelson and O. Zachar, Phys. Rev. B 56,
  6120 (1997); V.J. Emery and S.A. Kivelson, J.
Phys. Chem. Sol 59, 1705 (1998);
V.J. Emery and Kivelson, Physica C 282, 174 (1997) 


\item R.S. Markiewicz, J. Phys. Chem. Sol. 58, 1179 (1997)
 
 
\item R.A. Klemm, \lq\lq Layered Superconductors\rq\rq, Oxford University Press (1997)


\item T. Dahm, D. Manske and L. Tewordt, Phys. Rev. B 55, 15274 (1997)


\item S. Onoda, M. Imada, J. Phys. Soc. Jpn (submitted); see 
cond-mat 9903030 (1999) 


\item M. Randeira $et$ $al.$ Phys. Rev. Lett. 69, 2001 (1992)


\item N. Trivedi and M. Randeira, Phys. Rev. Lett. 75, 312 (1995)


\item A.A. Varlamov, G. Balestrino, E. Milani
and D.V. Livanov, \lq\lq The role of Density of States
Fluctuations in the Normal State Properties of High $T_c$
Superconductors\rq\rq, Adv. Phys. (1999)


\item C. Castellani, C. Di Castro and M. Grilli, Z. Phys. B 103,
137 (1997)


\item M. Randeira cond-mat/9710223 (1997) 


\item O. Zachar cond-mat/9902130 (1999)


\item P. Carretta,
D.V. Livanov, A. Rigamonti and A.A. Varlamov,
 Phys. Rev. B, 54, R9682
(1996)


\item M. Randeira and A.A. Varlamov, Phys. Rev B 50, 10401 (1994)


\bibitem{Vesna} V.F. Mitrovi\'c 
$et$ $al.$ Phys. Rev. Lett., 82, 2784
(1999)


\item H.N. Bachman $et$ $al.$, Phys. Rev. Lett. (1999) to be published




\item M. Eschrig, D. Rainer, J.A. Sauls,
Phys. Rev. B, 59, 12095 (1999)


\item
K. Gorny 
$et$ $al.$
Phys. Rev. Lett. 82, 177 (1999)


\item P. Carretta,
A. Lascialfari, A. Rigamonti, A. Rosso and A.A. Varlamov,
 Int. J. Mod. Phys. B (1999) to be
published


\item  A.I. Buzdin, A.A. Varlamov,
Phys. Rev. B, 58, 14195 (1998)


\item V. Dorin, R. Klemm, A.A. Varlamov, A. Buzdin and 
D. Livanov, Phys. Rev. B 48, 12951 (1993)
 
 
\item V. Pasler $et$ $al.$, Phys. Rev. Lett. 81, 1094 (1998)



\item A. Junod  
$et$ $al.$ Physica C 294, 115 (1998)


\item P. Carretta, A. Lascialfari, A. Rigamonti, A. Rosso and A.A.
Varlamov, Phys. Rev. B (1999) submitted






%


\end{enumerate}


\newpage
\section*{Captions for figures}
Fig. 1: Theoretical behaviours for the DOS contribution
to the nuclear spin-lattice relaxation rate
(normalized to the value in the absence of SF, $W^0$),
according to 
 Eqs. (3) and (5), in the text, as a function of temperature for
different values of the field (a), and as a function of the 
reduced magnetic field $\beta=2B/B_{c2}(0)$ for fixed values
of the temperature (b). The curves have been derived in
correspondence to choices of the
 upper critical field $B_{c2}(0)\sim 120$ T,
the Fermi energy $E_F=3500$ K$_B$ and the single-particle
collision time $\tau=2\cdot 10^{-14} $ s.
\\


Fig. 2: Theoretical 
behaviours for the MT contribution
to the nuclear spin-lattice
relaxation rate, according to
Eq. (1), in the case of strong pair-breaking
($\gamma_\varphi \simeq 0.3$).
Part (a) of Fig. 2 shows $W^{MT}(\beta, \varepsilon)/W^0$
as a function of temperature (for different values of the field)
while part (b) reports it as a function of the reduced field,
$\beta=2B/B_{c2}(0)$, in correspondence to fixed 
values of the temperature.
The choice of parameters is the same
 as in Fig. 1.
\\


Fig. 3: Comparison between the experimental results,
obtained at $T\simeq 95$ K ($\Box$ Mitrovi\'c $et$ $al.^{18}$; 
\ $\Diamond$ Gorny $et$ $al.^{21}$; \ 
$\Box$ da Carretta $et$ $al.^{16}$; \ 
$\times$ da Carretta $et$ $al.$ (data unpublished))
and the theoretical predictions for
the DOS term as a function of the reduced field
 in correspondence to the usual choice
of parameters (Figs. 1 and 2).
The experimental data by 
Gorny $et$ $al.^{22}$ ($\Diamond$) and by Carretta $et$ $al.$
($\times$) (unpublished) have been reported in correspondence
to the value $W^{DOS}(0,\varepsilon)\simeq
-0.16W^0$, in order to analyze
their possible field dependence.



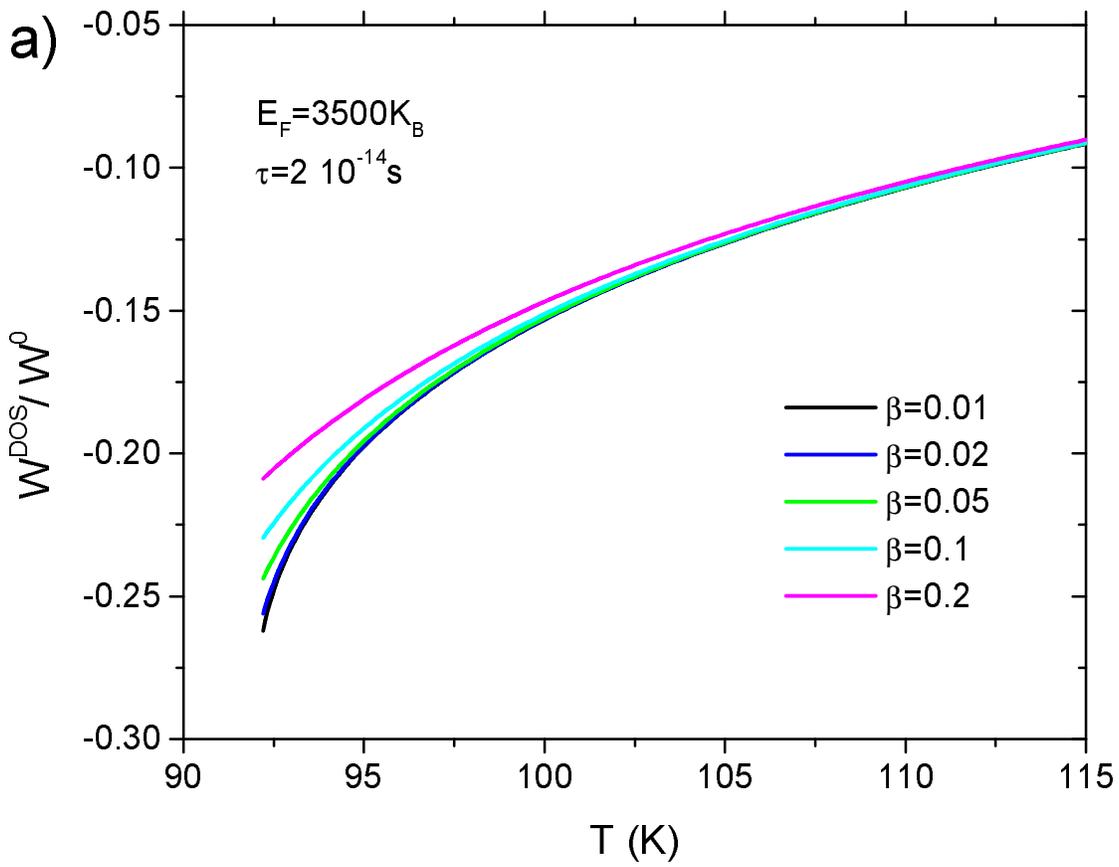
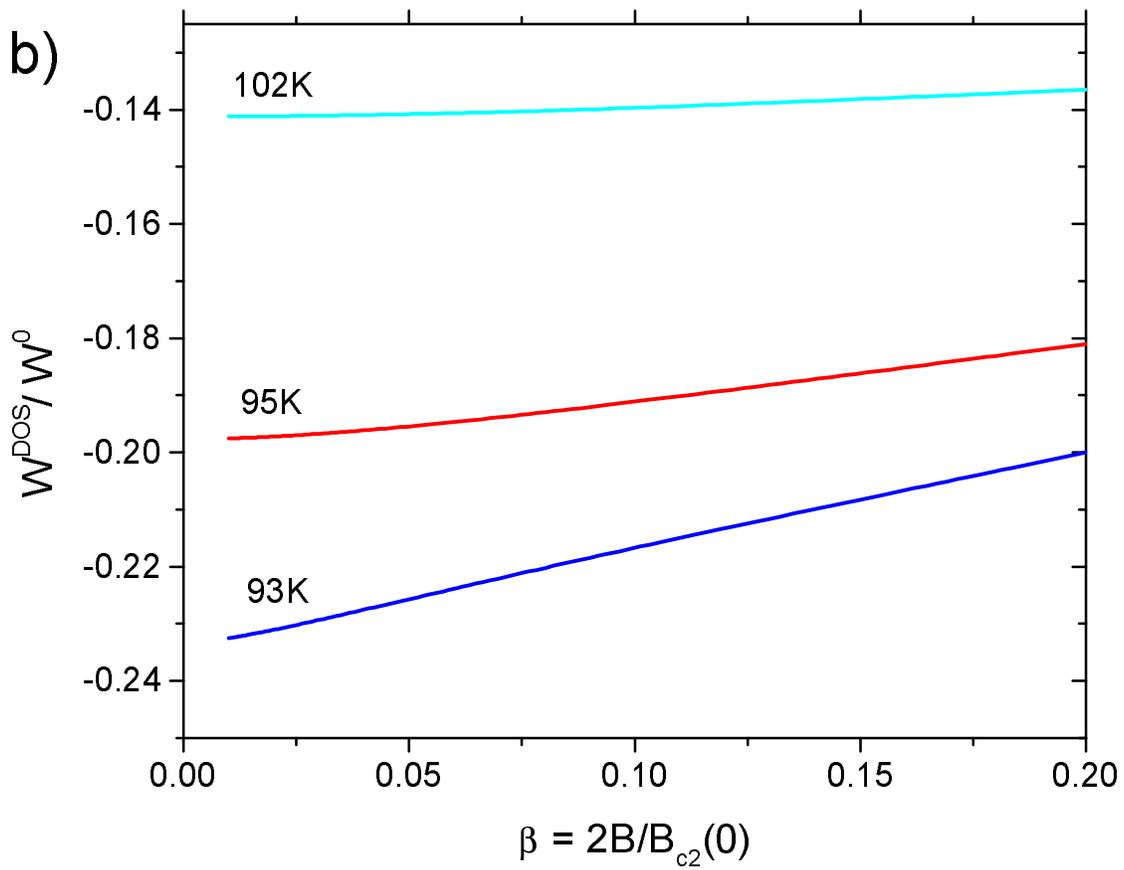
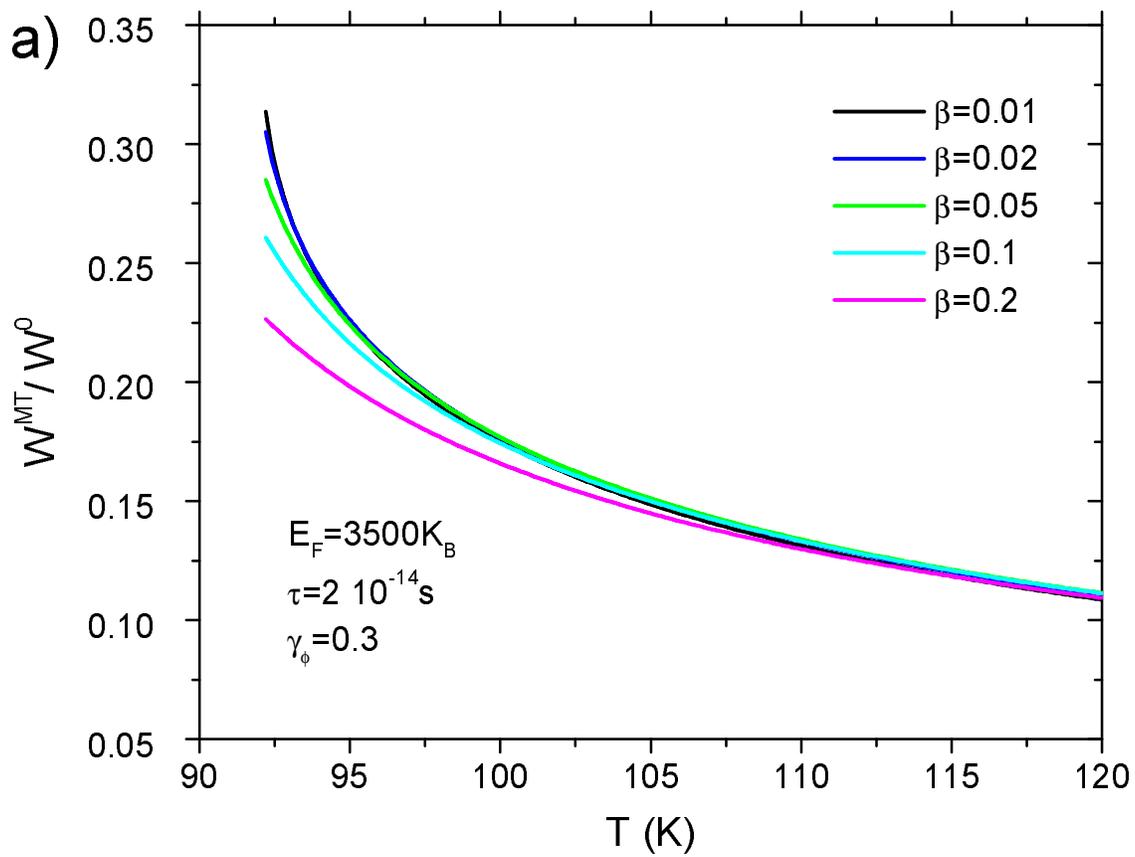
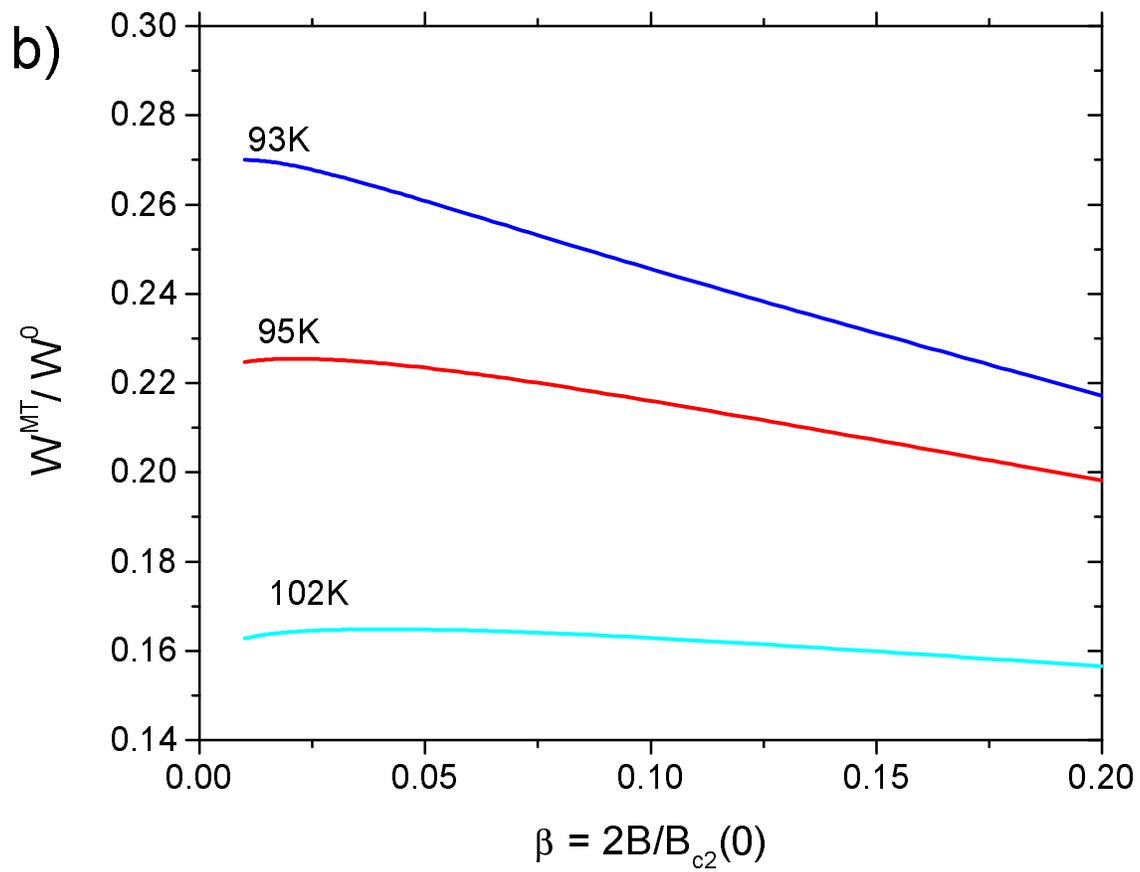
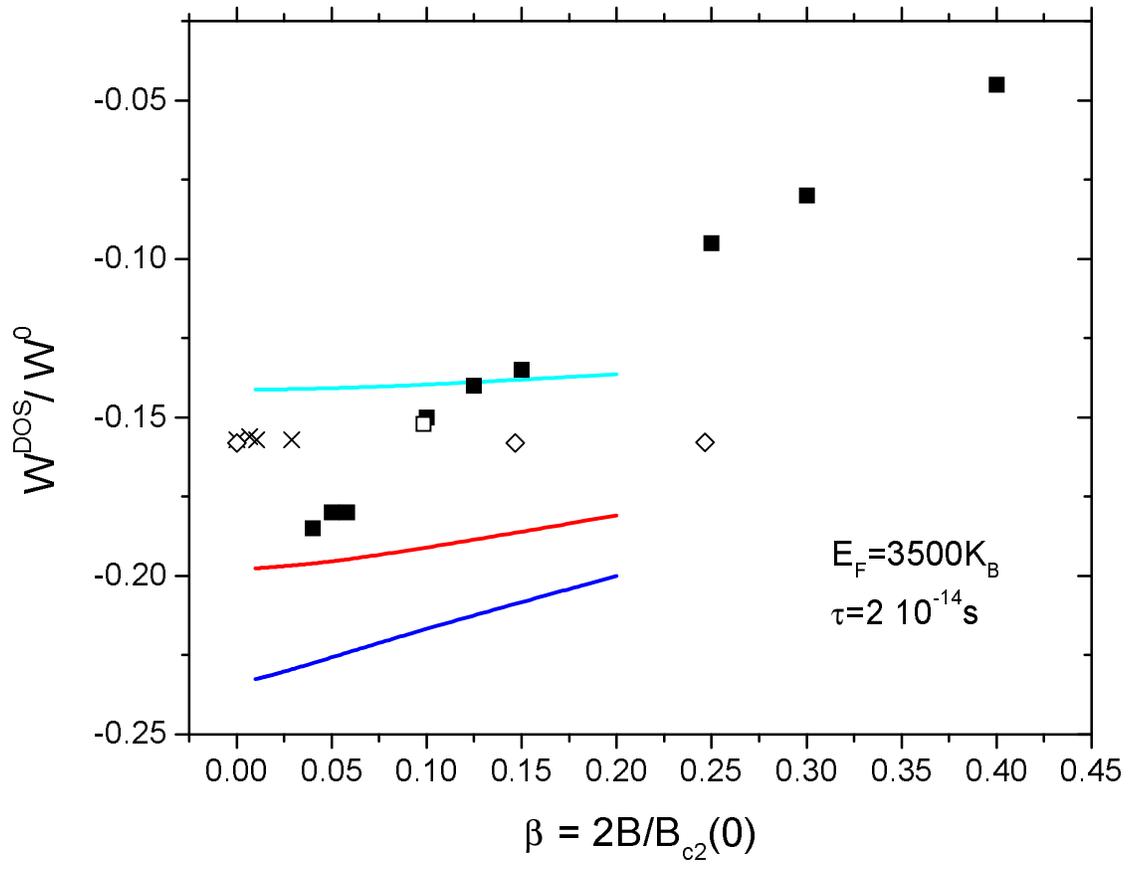
\end{document}